\documentclass[twocolumn,PRB,aps,psfig,preprintnumbers,superscriptaddress]{revtex4}

\usepackage{graphicx}
\usepackage{epstopdf}
\usepackage{float}
\usepackage{color}
\usepackage{amsmath}

\mathchardef\mhyphen="2D

\begin{document}

\title{Interrelationships between nematicity, antiferromagnetic spin fluctuations and superconductivity: Role of hotspots in FeSe$_{1-x}$S$_{x}$ revealed by  high pressure $^{77}$Se NMR study}

% Interrelationships between nematicity, antiferromagnetic fluctuations and superconductivity in FeSe$_{1-x}$S$_{x}$ under pressure;  }
%\chapter
% \title{Tuning Antiferromagnetic Correlations in FeSe$_{1-x}$S$_x$ under Pressure at the BCS-BEC Crossover}
%TUNING ANTIFERROMAGNETIC CORRELATIONS IN \\ FeSe$_{1-x}$S$_x$ UNDER PRESSURE AT THE BCS-BEC CROSSOVER}
%Tuning Antiferromagnetic Correlations in FeSe$_{1-x}$S$_x$ Under Pressure at the BCS-BEC Crossover

\author{K. Rana}
\affiliation{Ames National Laboratory, U.S. DOE, Ames, Iowa 50011, USA}
\affiliation{Department of Physics and Astronomy, Iowa State University, Ames, Iowa 50011, USA}
\author{D. V. Ambika}
\affiliation{Ames National Laboratory, U.S. DOE, Ames, Iowa 50011, USA}
\affiliation{Department of Physics and Astronomy, Iowa State University, Ames, Iowa 50011, USA}
\author{S. L. Bud'ko}
\affiliation{Ames National Laboratory, U.S. DOE, Ames, Iowa 50011, USA}
\affiliation{Department of Physics and Astronomy, Iowa State University, Ames, Iowa 50011, USA}
\author{A. E. B\"ohmer}
\affiliation{Ames National Laboratory, U.S. DOE, Ames, Iowa 50011, USA}
\affiliation{Department of Physics and Astronomy, Iowa State University, Ames, Iowa 50011, USA}
\affiliation{Institut f\"ur Experimentalphysik IV, Ruhr-Universit\"at Bochum, 44801 Bochum, Germany}
\author{P. C. Canfield}
\affiliation{Ames National Laboratory, U.S. DOE, Ames, Iowa 50011, USA}
\affiliation{Department of Physics and Astronomy, Iowa State University, Ames, Iowa 50011, USA}
\author{Y. Furukawa}
\affiliation{Ames National Laboratory, U.S. DOE, Ames, Iowa 50011, USA}
\affiliation{Department of Physics and Astronomy, Iowa State University, Ames, Iowa 50011, USA}
\date{\today}

\begin{abstract}
   The sulfur-substituted FeSe, FeSe$_{1-x}$S$_{x} $, is one of the unique systems that provides an independent tunability of nematicity, antiferromagnetism and superconductivity under pressure ($p$).
   Recently Rana {\it et al}. [Phys. Rev. B {\bf 101}, 180503(R) (2020)] reported, from $^{77}$Se nuclear magnetic resonance (NMR) measurements on  FeSe$_{0.91}$S$_{0.09}$ under pressure, that there exists a clear role of nematicity on the relationship between antiferromagnetic (AFM) spin fluctuations and superconducting transition temperature ($T_{\rm c}$)  where the AFM spin fluctuations are more effective in enhancing $T_{\rm c}$ in the absence of nematicity than  with nematicity. 
    Motivated by the work, we carried out  $^{77}$Se NMR measurements  on  FeSe$_{1-x}$S$_{x}$ with $x$~=~ 0.15 and 0.29 under pressure up to 2.10 GPa to investigate the relationship in a wide range of $x$ in the FeSe$_{1-x}$S$_x$ system. 
   Based on the new results together with the previously reported data for $x$~=~0 [P. Wiecki {\it et al.,} Phys. Rev. B {\bf 96}, 180502(R) (2017)] and 0.09  [K. Rana {\it et al.} Phys. Rev. B {\bf 101}, 180503(R) (2020)], we established a $p$ - $x$ - temperature ($T$) phase diagram exhibiting the evolution of AFM spin fluctuations. 
    From the systematic analysis of the NMR data, we found that  the superconducting (SC) state in nematic state arises from a non Fermi liquid state with strong stripe-type AFM spin fluctuations while the SC state without nematicity comes from a Fermi liquid state with mild stripe-type AFM spin fluctuations. 
    Furthermore, we show that the previously reported impact of nematicity on the relationship between AFM fluctuations and superconductivity holds throughout  the wide range of $x$  from $x$ = 0 to 0.29 in FeSe$_{1-x}$S$_{x}$ under pressure.  
   We discuss the origin of the role of nematicity  in terms of  the different numbers of hotspots on Fermi surfaces with and without nematicity.

\end{abstract}

\maketitle

\section{Introduction}

    The interplay between magnetic fluctuations, electronic nematicity and the unconventional nature of superconductivity has received wide interest after the discovery of high $T_{\rm c}$ superconductivity in iron pnictides \cite{Kamihara2008}.
%    Superconducting (SC) pairing in unconventional superconductors such as heavy-fermion systems \cite{Curro2005,Hattori2012,Scalipino2012}, high-$T_{\rm c}$ cuprates \red {[add refs]} and Fe-based superconductors (FeSCs)\cite{Scalipino2012,Bertel2016} have often been discussed in terms of the critical fluctuations of an ordered phase found in close proximity of the SC state.
     In most  Fe-based superconductors (FeSCs), superconductivity appears close to the quantum phase transitions of two long-range orders: the nematic order, which is an electronically driven structural transition from high-temperature tetragonal (HTT) with C4 symmetry to low-temperature orthorhombic (LTO) having C2 symmetry, and the antiferromagnetic (AFM) order with spontaneously oriented Fe-3$d$ spins characterized by stripe-type spin structure \cite{Scalipino2012,Johnston2010,Canfield2010,Stewart2011}. %characterized by a wave vector [${\bf q}$ = ($\pi$,0) or (0,$\pi$)]    \cite{Scalipino2012,Johnston2010,Canfield2010,Stewart2011}. 
    In those systems, the nematic transition temperature ($T_{\rm s}$) is usually at or just above the transition temperature for the AFM state, ($T_{\rm N}$), and superconductivity emerges upon the simultaneous suppression of both the  nematic and the AFM transitions by carrier doping and/or the application of pressure ($p$).  
   These results clearly indicate a close relationship between AFM and nematic phases, however, the individual contribution to superconductivity from these two phases is difficult to separate due to the close proximity of the two phases.
%    While AFM fluctuations have been thought as the primary candidate responsible for Cooper pairing in unconventional superconductors \red{[add refs.]},  nematic fluctuations are also considered to contribute to the SC pairing mechanism \cite{Kuo2016,Lederer2015,Lederer2017}. 

    In this respect,  the S-substituted FeSe system,  FeSe$_{1-x}$S$_x$, provides a suitable platform to investigate  the individual contribution of nematic and AFM fluctuations to superconductivity.
%  due to the existence of a nematic order independent of the AFM order. 
    FeSe$_{1-x}$S$_x$ has the simplest of crystal structures among the Fe-based superconductors, with quasi-two dimensional Fe(Se,S) layers in the $ab$ plane, stacked along the $c$ axis. 
    At $x$ = 0,  FeSe undergoes a nematic phase transition at $T_{\rm s}$ $\sim$ 90 K followed by a superconducting (SC) transition at $T_{\rm c}\sim$~8.5~K without AFM ordering at ambient pressure \cite{Bohmer2018,Hsu2008,McQueen2009,Baek2015}.
   This allows the study of magnetic fluctuations inside the nematic order and its relationship with superconductivity \cite{Imai2009}. 
   The nematic phase in FeSe can be suppressed by pressure  application, with $T_{\rm s}$  decreasing down to 32~K  at $p$~=~1.5~GPa \cite{Wiecki2018, Wang2017}.
   $T_{\rm c}$ shows a complex multi-domed structure with $p$, reaching a maximum $T_{\rm c}$ $\sim$ 37~K at $p\sim$~6~GPa \cite{Mizuguchi2008,Margadonna2009,Medvedev2009}. 
  At the same time,  an AFM ordered state appears above $p$~=~0.8~GPa \cite{Terashima2015,Bendele2010},  and $T_{\rm s}$ merges with $T_{\rm N}$ above $p$~=~1.7~GPa \cite{Kothapalli2016, Bohmer2019, Gati2019}, limiting the range for studying the effects of nematicity on superconductivity without AFM state.

   The  S-substitutions for Se in  FeSe$_{1-x}$S$_{x}$ is also a well known way to control the nematic phase where $T_{\rm s}$ is suppressed to zero at the critical $x$ value, $x_{\rm c}\sim$~0.17, with increasing $x$.
  $T_{\rm c}$ first  increases up to 10 K around $x$ = 0.09 making a maximum  and then decreases gradually at higher $x$ close to $x_{\rm c}$ \cite{Watson2015, Wiecki2018,Reiss2017,Abdel2015}.
     Although AFM order is not seen at ambient pressure in the FeSe$_{1-x}$S$_{x} $ system, an enhancement of AFM fluctuations was found at $x$  $\sim$ 0.09 by nuclear magnetic resonance (NMR) measurements \cite{Wiecki2018}. 
   The NMR study also suggested that AFM fluctuations are important in determining the superconductivity in FeSe$_{1-x}$S$_{x} $ even in the vicinity of a nematic quantum phase transition (QPT) at $x_{\rm c}$ $\sim$ 0.17. 
    On the other hand, recent spectroscopic-imaging scanning tunneling microscopy  \cite{Hanaguri2018}, thermal conductivity and specific heat \cite{Sato2018} measurements  show an abrupt change in $T_{\rm c}$ around $x_{\rm c}$ \cite{Mizukami2021} and the considerable change in the size and anisotropy of the SC gap is observed  at the nematic QPT, implying different SC states inside and outside  nematic states  \cite{Coldea2021}.  
    These results may indicate that the nematicity also plays an important role in the SC states of FeSe$_{1-x}$S$_{x} $.

    In fact, nematicity has been reported to affect superconductivity by changing the symmetry \cite{Fernandes2012,Fernandes2014} and magnitude \cite{Wiecki2018,Tanatar2016} of AFM spin fluctuations.  
    AFM spin fluctuations in FeSCs are characterized by the in-plane stripe-type wave vectors: $\bf{Q_{\rm 1}}=(\pi,0)$ and/or $\bf{Q_{\rm 2}}=(0,\pi)$ (using the single-iron Brillouin zone notation)  \cite{Paglione2010,Dai2015}. 
    In the HTT C4  symmetric state where nematicity is absent, the AFM spin fluctuations may originate from both $\pm \bf{Q_{\rm 1}}$ and $\pm \bf{Q_{\rm 2}}$ wave vectors. 
    In contrast, in the C2 symmetric state when nematicity is present, only one of the two wave vectors, either $\pm \bf{Q_{\rm 1}}$ or $\pm \bf{Q_{\rm 2}}$, will contribute to AFM spin fluctuations, due to the breaking of symmetry equivalency \cite{Fernandes2012,Fernandes2014}. 
   Recent $^{77}$Se NMR studies on FeSe$_{1-x}$S$_x$  under pressure actually showed the large change in the magnitude of AFM spin fluctuations with and without nematicity \cite{Rana2020,Kuwayama2021,Kuwayama2019,Kuwayama2020}. 
    In addition, from $^{77}$Se NMR studies on FeSe$_{1-x}$S$_x$ with $x$ = 0.09 under pressure,  the impact of nematicity on the relationship between AFM spin fluctuations and $T_{\rm c}$ has been proposed,  where the AFM spin fluctuations are more effective in enhancing superconductivity in the absence of nematicity as compared to when it is present \cite{Rana2020,Rana2022}.
    On the other hand, as described above, the importance of nematicity for the appearance in superconductivity has been discussed in  FeSCs \cite{Kontani2010,Shibauchi2020,Yang2015,Kuo2016,Lederer2015,Lederer2017}.  
    Therefore it is important to investigate the relationships between AFM spin fluctuations, nematicity, and superconductivity in a wide range of $x$ in FeSe$_{1-x}$S$_x$.

%,  although  $T_{\rm c}$ is roughly proportional to AFM spin fluctuations in the presence or absence of nematic order. 

   In this paper, we carried out  $^{77}$Se NMR studies under pressure up to 2.1 GPa for  $x = 0.15$ and $0.29$ compounds to investigate the universality of the role of nematicity on the relationship for a wide range of $x$ in FeSe$_{1-x}$S$_{x}$. 
   From the analysis of the NMR data for the two compounds together with the previous data for $x = 0$ \cite{Wiecki2017} and $x = 0.09$ \cite{Rana2020} systems under $p$, we report that  the SC state in nematic state  arises from a non-Fermi liquid (nFL) with strong AFM spin fluctuations  while the SC state without nematicity  arises from a Fermi liquid (FL) with moderate AFM spin fluctuations  in these systems.
   Furthermore, we confirm that nematicity has a clear impact on the relationship between AFM spin fluctuations and superconductivity throughout the FeSe$_{1-x}$S$_{x} $ system under $p$ as previously suggested \cite{Rana2020}. 
   The AFM spin fluctuations in the C4 phase were found to be more effective in enhancing $T_{\rm c}$ compared to AFM spin fluctuations in the C2 phase by a factor of $\sim 7 \pm 2$. We provide a possible explanation for this observation in terms of the higher total number of hotspots in the Fermi surface that meet the nesting condition in the C4 symmetric state, in comparison with the case of the C2 symmetric state by a factor of 8. 
   Finally it suggested that AFM correlation length ($\xi_{\rm AFM}$) in the C2 states is generally longer than in the C4 phase which could be related to the different nature of SC states with and without nematicity. 
% continuously decreases as the system moves from  SC$_{\rm {C2}}$ to SC$_{\rm {C4}}$ through out the $x\mhyphen p$ phase diagram.

\section{Experimental Details}
    NMR measurements of $^{77}$Se nuclei ($I$=1/2, $\gamma_{\rm N}/2\pi = 8.118 $MHz/T) were carried out using a laboratory-built phase-coherent spin-echo pulse spectrometer. 
  The measurements were carried out under a fixed magnetic external field of $H$ = 7.4089 T applied either along the $c$ axis or the tetragonal [110] direction in the $ab$ plane. 
   The single crystals with sulfur contents $x = 0.15$ and 0.29 in FeSe$_{1-x}$S$_{x} $ were grown using chemical vapor transport as outlined in Refs. \cite{Wiecki2018,Bohmer2016}. 
    Multiple small single crystals were oriented and stacked with spacers inside the NMR coil. 
    A NiCrAl/CuBe piston-cylinder was used for application of $p$ up till $\sim$ 2 GPa. 
   Daphne 7373 was used as the $p$ mediating fluid,  and the nuclear quadrupolar resonance (NQR) of Cu in Cu$_2$O at 77 K was used for $p$ calibration \cite{Fukazawa2007,Reyes1992}.  
 %  $^{77}$Se NMR spectra and spin-lattice relaxation time ($T_1$) were measured under a fixed magnetic field of 7.4089 T. 
   Spectra were obtained from spin-echo signals by using fast Fourier transform. 
   $T_1$ was measured with a saturation-recovery method and    $1/T_1$ at each $T$ was determined by fitting the nuclear magnetization $M$ versus time $t$ using the single  exponential function $1-M(t)/M(\infty) =  e^ {-t/T_{1}}$, where $M(t)$ and $M(\infty)$ are the nuclear magnetization at $t$ after saturation and the equilibrium nuclear magnetization at $t$ $\rightarrow$ $\infty$, respectively.

\begin{figure*}[htb!]
\includegraphics[width=2\columnwidth]{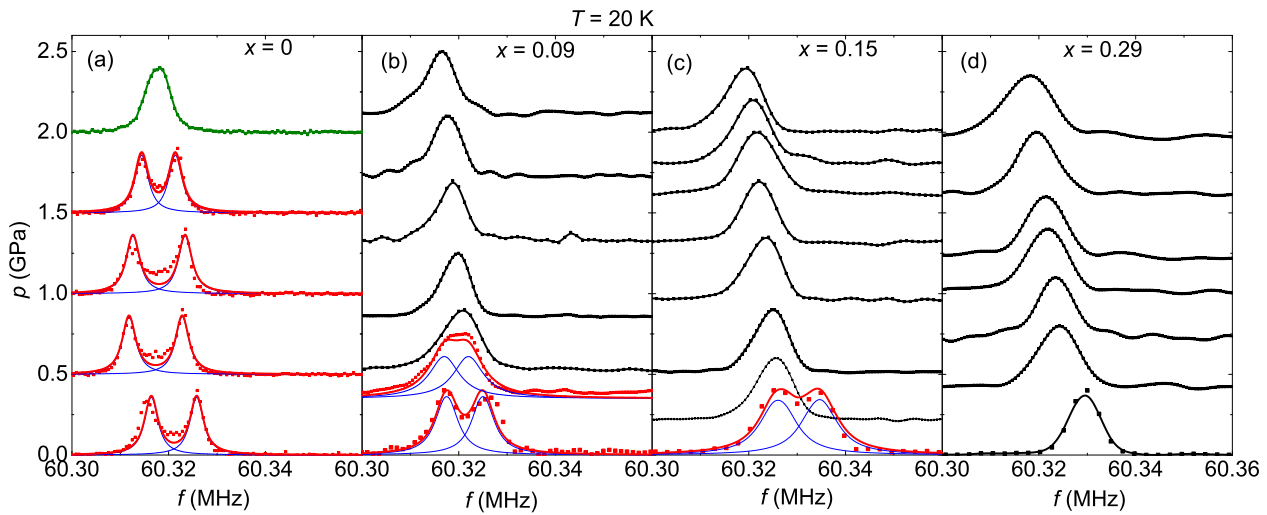} 
\caption[Pressure and $x$ dependence of $^{77}$Se NMR spectra of FeSe$_{1-x}$S$_{x} $ at the temperature of 20 K.] {Pressure ($p$) and $x$ dependence of $^{77}$Se NMR spectra of FeSe$_{1-x}$S$_{x} $ at a temperature of 20 K under the magnetic field of 7.4089 T applied along the tetragonal [110] direction in the $ab$ plane. 
   The spectra for (a) $x = 0$ was taken from Ref. \cite{Wiecki2017} and (b) $x = 0.09$ from Ref. \cite{Rana2020} whereas the spectra for FeSe$_{1-x}$S$_{x} $ at ambient pressure were taken from Ref. \cite{Wiecki2018}. 
    Red points represent the spectra in the nematic state which are fitted as the sum (red lines) of two Lorentzian peaks (blue lines).  Green line in (a) is the spectrum in antiferromagnetic state. Single peak spectra in the HTT phase are shown with black lines.}  
\label{fig:Fig1}
\end{figure*}

\section{Results}

\subsection{$^{77}$Se NMR Spectrum and Knight shift}

% \begin{figure*}%[tb]
% \includegraphics[width=2\columnwidth]{Fig3.eps} 
% \caption[Temperature dependence of $^{77}$Se NMR Knight shift ($K$) for $x = 0.29$ at various measured pressures.]{Temperature  dependence of $^{77}$Se NMR Knight shift ($K$) for $x = 0.29$ at various measured pressures under a magnetic field ($H$) of 7.4089 T, with (a) $H \parallel ab$ ($K_{\rm {ab}}$) and (b) $H \parallel c$ ($K_{\rm {ab}}$) directions. Here pressure is indicated as an implicit parameter. Corresponding insets present the low $T$ dependence of (a) $K_{\rm {ab}}$ and (b) $K_{\rm {c}}$ at selected pressures where straight lines indicate the saturation values while lowering $T$. }
% \label{fig:Fig3}
% \end{figure*} 

    Figures \ref{fig:Fig1}(a)-\ref{fig:Fig1}(d) show the  $^{77}$Se NMR spectra measured at $T$ = 20 K under various pressures ($p$ = 0 - 2.10~GPa)  for $x$ = 0.15 and 0.29, together with those for $x$ = 0 and 0.09 reported previously \cite{Wiecki2017, Rana2020}. 
   Here we applied magnetic field along [110] direction in the HTT phase ($H\parallel ab$). 
  As in the case of $x$ = 0 ($T_{\rm s}$ = 90 K) and 0.09 ($T_{\rm s}$ = 65 K) at ambient $p$, the two split lines of the NMR spectrum (shown in red) are observed in nematic state for $x$ = 0.15  ($T_{\rm s} = 35$ K at ambient $p$) as shown at the bottom of Fig.  \ref{fig:Fig1}(c). 
   Those two peak structures are due to the twinned nematic domains \cite{Baek2015,Bohmer2015,Wiecki2017}. 
As can be seen in  Fig. \ref{fig:Fig1}(c) for $x$ = 0.15,  the splitting in the NMR spectra is not seen at $p$ =~0.2 GPa and higher (even down to 1.7~K~at 0.2~GPa, as shown in the supplementary \cite{Sup}).
This indicates a very small critical pressure of  $p_{\rm c}$~$\sim$~ 0.2~GPa for nematic QPT at $x$~ =~ 0.15.  
    In the case of $x = 0.29$ ($>$  $x_c$ $\sim$ 0.17), no split of the NMR spectra is observed from ambient pressure up to 1.95~GPa as shown in Fig. \ref{fig:Fig1}(d), showing no-nematic state in the compound.   
   NMR spectra for nonzero $x$ values are broader  than those for $x = 0$ due to the introduction of inhomogeneity with sulfur substitution of selenium. 
  %     On the other hand, no clear splitting of the line can be observed in $x$ = 0.29 under any $p$, indicating no nematic phase.  

\begin{figure*}[tb]
\includegraphics[width=2\columnwidth]{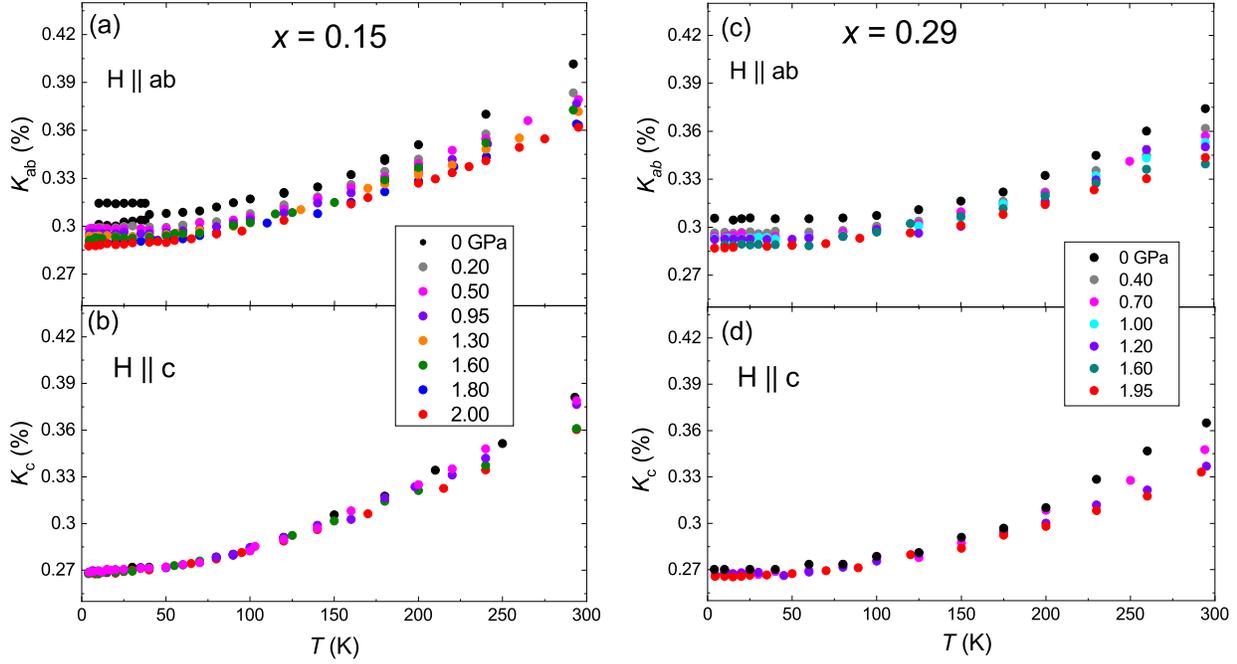} 
\caption{Temperature and pressure dependencies of $^{77}$Se NMR Knight shift ($K$) at various measured pressures under a magnetic field ($H$) of 7.4089 T for  (a) $K_{\rm {ab}}$ and (b) $K_{\rm {c}}$ for $x = 0.15$. (c) $K_{\rm ab}$ and (d) $K_{\rm c}$ for $x$ = 0.29. }
\label{fig:Fig2}
\end{figure*}

    Figures \ref{fig:Fig2}(a) and \ref{fig:Fig2}(b) show  the $T$ dependences of $K$ for $H\parallel ab$ ($K_{\rm {ab}}$) and  for $H \parallel c$ ($K_{\rm c}$)  of $x = 0.15$ at various pressures from ambient up to 2.0 GPa. 
   With decreasing $T$, both $K_{\rm ab}$ and $K_{\rm c}$ decrease and become nearly constant below $\sim$50 K. 
   The values of $K_{\rm ab}$ decrease slightly with increasing $p$ while $K_{\rm c}$ shows less $p$ dependence. 
   In particular,  no clear $p$ dependence is observed in $K_{\rm c}$ at low $T$ below $\sim$ 150 K within our experimental uncertainty.
    Similar $T$ and $p$ dependences of $K_{\rm ab}$ and $K_{\rm c}$ are observed for $x = 0.29$ as shown in Figs. \ref{fig:Fig2}(c) and \ref{fig:Fig2}(d). 

\begin{figure}[tb]
\centering
\includegraphics[width=1\columnwidth]{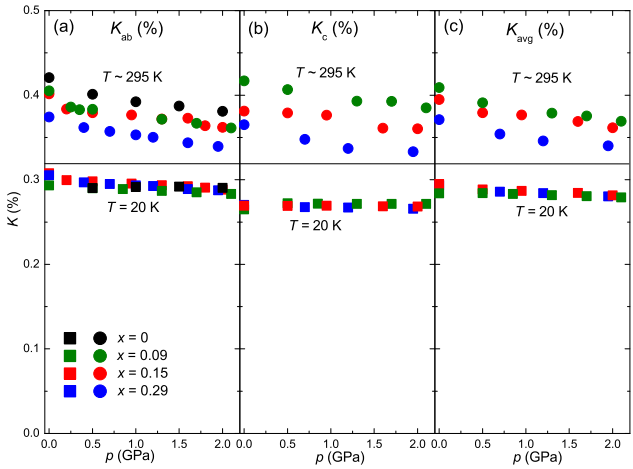} 
\caption[Pressure and $x$  dependences of the Knight shift in FeSe$_{1-x}$S$_{x}$.]{Pressure and $x$ dependences of the Knight shift ($K$) under a magnetic field ($H$) of 7.4089~T for  (a) $H\parallel ab$, (b) $H \parallel c$,  and (c) the average value of Knight shift ($K_{\rm {avg}} \equiv \frac{2K_{\rm {ab}}+K_{\rm c}}{3}$) at different temperatures at 295 K (top) and 20 K (bottom).  The data for $x$ = 0 and 0.09 are from Ref. \cite{Wiecki2017} and Ref. \cite{Rana2020} respectively and the $K_c$ values under $p$ for $x$ = 0 are missing due to the unavailability of the data.} 
\label{fig:Fig4}
\end{figure}

   To see the $x$ and $p$ dependences of $K$ at high and low temperatures more clearly, we plot the  $K_{\rm {ab}}$ and $K_{\rm {c}}$ measured at 295 K (top) and at 20 K (bottom) in Fig. \ref{fig:Fig4}(a) and \ref{fig:Fig4}(b), respectively.    
  We also plot the average values of Knight shift ($K_{\rm {avg}} \equiv \frac{2K_{\rm {ab}}+K_{\rm c}}{3}$) in Fig. \ref{fig:Fig4}(c). 
   Here the values of $K$ for $x = 0$ and 0.09 were taken from Refs. \cite{Wiecki2017} and \cite{Rana2020}, respectively. 
   The average values of $K_{\rm {ab}}$ for the two peaks were used when nematic state was present. 
    In general, $K$ consists of the $T$-independent orbital component ($K_0$) and the $T$-dependent spin component ($K_{\rm {spin}}$) which is proportional to static uniform magnetic susceptibility and thus to the density of states at the Fermi energy $N(E_{\rm F})$. 
   As we show later in Sec. IV, $K_0$ is nearly independent of $x$ and $p$.   
   Therefore the $x$ and $p$ dependences of $K_{\rm ab}$, $K_{\rm c}$ and $K_{\rm ave}$ can be attributed to the spin part of $K$.  
   Given those dependences, the $N(E_{\rm F})$ in FeSe$_{1-x}$S$_{x} $ system has a slight suppression at 295 K with increasing $x$ and/or $p$ up to $\sim$ 2.0 GPa. 
   In contrast,  the $N(E_{\rm F})$ is nearly independent of $x$  at low temperatures with a tiny decrease with $p$.
We note that the change in $N(E_{\rm F})$ due to a Lifshitz transition under pressure suggested in the quantum oscillations measurements for $x$~=~0.11 \cite{Reiss2019} and inferred from NMR measurements for $x$~=~0.12 \cite{Kuwayama2021} was not detected in the $p$ dependence of $K$ values within our experimental uncertainty in the measured FeSe$_{1-x}$S$_{x} $ systems.
   It is interesting to point out that $T_{\rm c}$ varies significantly with $x$ and $p$ even though the $N(E_{\rm F})$ is nearly independent of those. 
      This is in contrast to conventional BCS superconductors, in which $N(E_{\rm F})$ generally correlates with $T_{\rm c}$.
     These results strongly indicate that AFM spin fluctuations play an important role in the appearance of SC in FeSe$_{1-x}$S$_{x}$, as will be discussed below.   
%    Recently, $^{77}$Se NMR measurements  on $x$ = 0.12 have suggested a $p$ induced Lifshitz transition based on a non-monotonic change in $K_{\rm {spin}}$ values \cite{Kuwayama2021}.
% where $p$ dependence of $K_{\rm {orb}}$ were carefully estimated from the clear decrease of $K$ at the  $T_{\rm c}$ with $H$ = 6.02 T. 
  % However,  we do not see a clear change in $K$ values in our data for $x$ = 0, 0.09 , 0.15 and 0.29  within our experimental uncertainty. 

%\begin{figure}[tb]
%\includegraphics[width=\columnwidth]{KS29pcHpc} 
%\caption{Temperature dependence of $^{77}$Se NMR Knight shift with H$||c$ for x=0.29 with pressure as an implicit parameter}
%\label{fig:Fig3b}
%\end{figure} 

% \begin{figure}[h!tb]

\subsection{$^{77}$Se Spin-lattice   Relaxation Rates  ($1/T_1$)}

\begin{figure*}[h!tb]
\includegraphics[width=2\columnwidth]{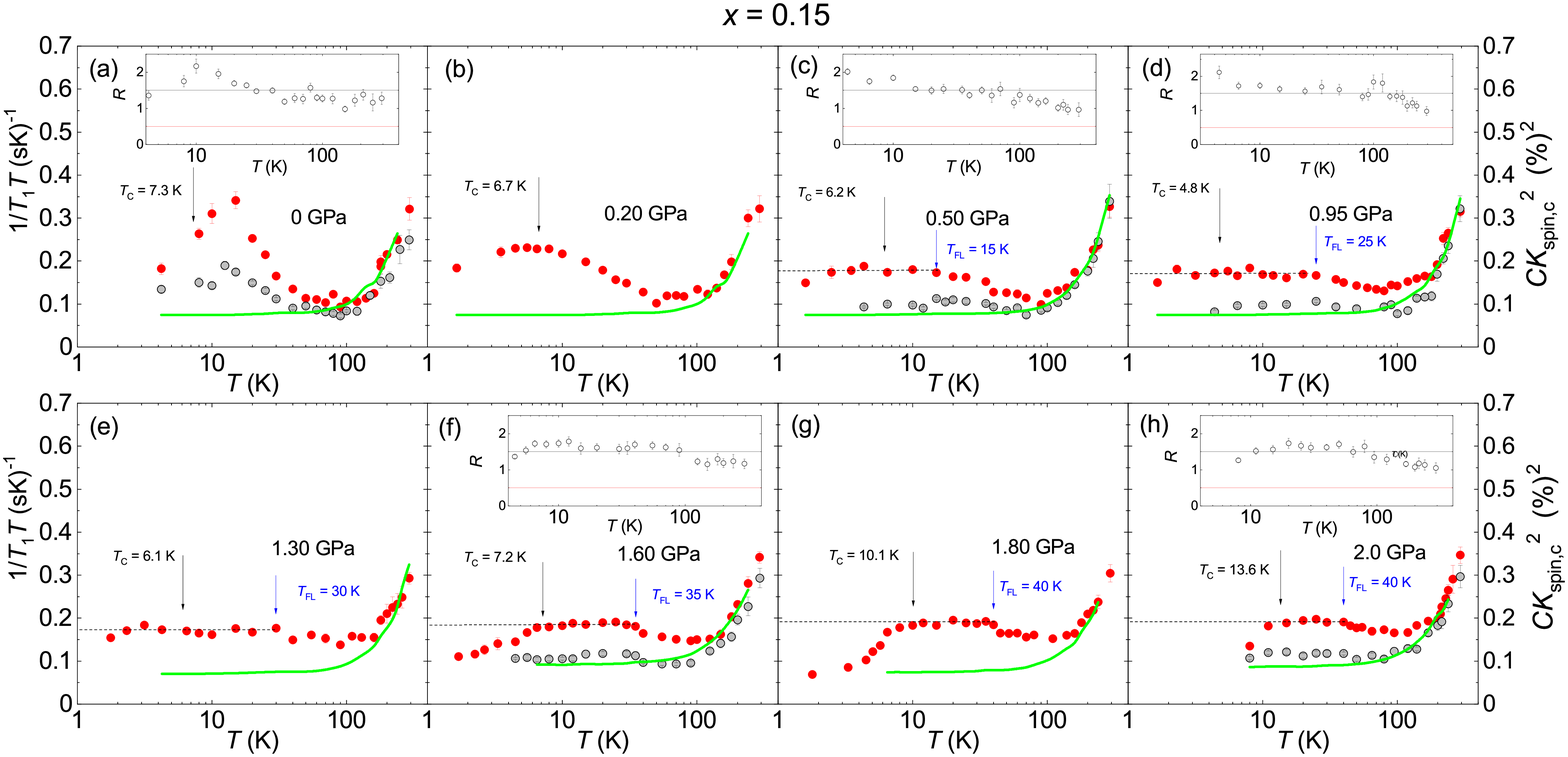} 
\caption{Temperature dependence of $^{77}$Se 1/$T_1T$ (left axes) for $x = 0.15$ at various pressures under a magnetic field ($H$)~=~7.4089 T, with $H \parallel ab$ (red circles) and $H \parallel c$ (gray circles). 
   The green line for each panel (right axes) shows the temperature dependence of $CK_{\rm spin,c}^2$ where $K_{\rm spin,c}$ is the spin part of $K$ estimated by subtracting the $T$ independent $K_0$ and $C$ is a scaling factor. 
   For all pressures measured, $K_0$ and $C$ are found to be constant, 0.175 \% and  8.5, respectively. 
    Downward black arrows show $T_{\rm c}$ under $H \parallel ab$ = 7.4089 T determined by the ${\it in~situ}$ ac susceptibility measurements \cite{Sup} and downward blue arrows show the $T$ below which $1/T_1T$ = constant (black dashed line) behavior is observed, defined as $T_{\rm FL}$. 
The inset in each panel  shows the $T$ dependence of  the ratio $R$ $\equiv$ $(1/T_1T)_{ab}$/$(1/T_1T)_{c}$. 
    The black and red  horizontal lines represent the expected values for stripe-type ($R$ = 1.5) and  N\'eel-type ($R$ = 0.5) antiferromagnetic fluctuations, respectively. }
\label{fig:Fig5}
\end{figure*}

% In particular, in the case of FeSe$_{1-x}$S$_{x} $ system, the presence of AFM spin fluctuations is evidenced at low temperatures around $\sim$ 100 K and much above $T_{\rm c}$where K becomes constant but $1/T_1T$ shows an enhancement. This is because only $1/T_1T$ is sensitive to $q\neq 0$ components of $\chi \prime$  unlike K. 

 %  We now show the dynamic properties of the FeSe$_{1-x}$S$_{x} $ system using $^{77}$Se spin-lattice  relaxation time ($T_1$). 
   Figures \ref{fig:Fig5}(a)-\ref{fig:Fig5}(h) show the $T$ dependence of $^{77}$Se $1/T_1T$ for $H \parallel ab$ [$(1/T_1T)_{\rm ab}$]  by red circles for the various pressures at $x = 0.15$. 
   In general, 1/$T_1T$ is related to the dynamic magnetic susceptibility as $1/T_1T\sim\gamma^{2}_{\rm N}k_{\rm B}\sum_{\mathbf{q}}|A(\mathbf{q})|^2\chi^{\prime\prime}(\mathbf{q}, \omega_{\rm N})/\omega_{\rm N}$, where $A(\mathbf{q})$ is the wave-vector $\mathbf{q}$ dependent form factor and $\chi^{\prime\prime}(\mathbf{q}, \omega_{\rm N})$ is the imaginary part of $\chi(\mathbf{q}, \omega_{\rm N})$ at the Larmor frequency $\omega_{\rm N}$  \cite{Moriya1963}. 
  Since $K$ measures the $\mathbf{q}=0$ uniform magnetic susceptibility, by comparing the $T$ dependencies between 1/$T_1T$ and $K$,  one can obtain information on the $T$ evolution of $\sum_{\mathbf{q}}\chi^{\prime\prime}(\mathbf{q}, \omega_{\rm N})$ with respect to that of $\chi^{\prime}(0, 0)$. 
   Above $\sim$ 100 K, $(1/T_1T)_{\rm ab}$ decreases monotonically with decreasing $T$ similar to $K_{\rm{ab}}$ and $K_{\rm{c}}$. 
   In contrast, different $T$ dependences below $\sim$ 100 K were observed: $K_{\rm{ab}}$ and $K_{\rm{c}}$ continue to decrease on lowering $T$ and show a nearly constant behavior below 50 K for $K_{ab}$ and 25 K for $K_c$  respectively, whereas $(1/T_1T)_{\rm ab}$ starts increasing   around $\sim$~100~K when decreasing $T$. 
   This deviation is attributed to the presence of AFM spin fluctuations as in $x = 0$ and 0.09 in  FeSe$_{1-x}$S$_{x}$ \cite{Wiecki2018,Rana2020}.  
  The decrease in $1/T_1T$ observed around or below $T_{\rm c}$ marked by black arrows is due to SC transition.
  The suppression of 1/$T_1T$ above $T_{\rm c}$ is also observed  [see Fig. \ref{fig:Fig5}(a)] whose origin has been discussed in terms of a possible SC fluctuation effect \cite{Kasahara2016} or pseudogap behavior \cite{Shi2018}. 
 
   The temperature dependence of $(1/T_1T)_{\rm ab}$  at low temperatures changes when $p$ is passed $p_{\rm c} \sim$ 0.2 GPa for $x$~=~0.15. 
%   The nature of AFM spin fluctuations gradually changes as $p$ is passed $p_{\rm c} \sim$ 0.2 GPa for $x$ = 0.15. 
  As shown in Fig. \ref{fig:Fig5}(a), at ambient $p$, the $(1/T_1T)_{\rm ab}$ at low $T$ shows a Curie-Weiss like enhancement which is  expected for  two-dimensional AFM spin fluctuations \cite{Moriya1963}. 
    On the other hand, for $p > p_{\rm c}$ shown in Figs. \ref{fig:Fig5}(c)-\ref{fig:Fig5}(h), $(1/T_1T)_{\rm ab}$ moderately increases below $\sim$ 100 K before exhibiting a constant behavior below a characteristic temperature $T_{\rm{ FL}}$ marked by blue arrows. Here we determined $T_{\rm{ FL}}$ from the $1/T_1T$ data for $H||ab$ since $1/T_1T$ are more sensitive to AFM fluctuations for $H||ab$ than $H||c$ \cite{Wiecki2018}.
    A similar temperature dependence of 1/$T_1T$ has been reported for $x = 0.09$ \cite{Rana2020} and   the state below $T_{\rm{ FL}}$ was regarded as a Fermi liquid (FL) state because it follows the Korringa behavior where both $1/T_1T$ and $K^2$ are constant, similar to the FL state in exchange enhanced metals \cite{Moriya1963,Narath1968}. 
%   $1/T_1T$ for $H \parallel ab$ starts to decrease while decreasing $T$ below $T_{\rm c}$ marked by black arrows. 
  It is important  to note that while the SC state with C4 symmetry  arises from a FL state with AFM spin fluctuations, the SC state with C2 symmetry  arises from the nematic state with AFM spin fluctuations that show a Curie-Weiss-like  behavior.

\begin{figure*}[h!bt]
\includegraphics[width=2\columnwidth]{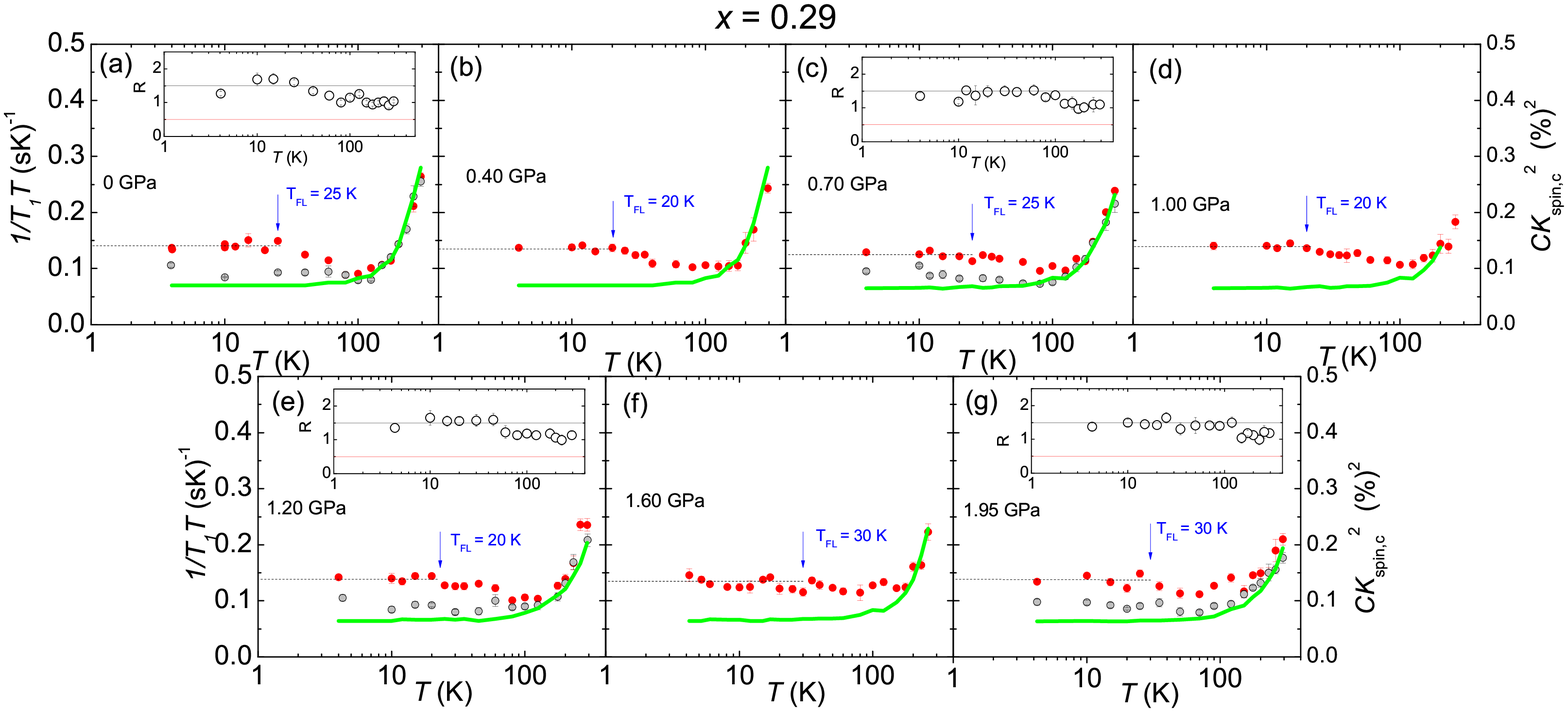} 
\caption{Temperature dependence of $^{77}$Se 1/$T_1T$ (left axes) for $x = 0.29$ at various pressures under a magnetic field ($H$) of 7.4089 T, with $H \parallel ab$ (red circles) and $H \parallel c$ (gray circles). 
   The green line for each panel (right axes) shows the temperature dependence of $CK_{\rm spin,c}^2$ where $K_{\rm spin,c}$ is the spin part of $K$ estimated by subtracting the $T$ independent $K_0$ and $C$ is a scaling factor. 
   For all pressures measured, $K_0$ and $C$ are found to be constant, 0.175 \% and  7.75, respectively. 
    Downward blue arrows show the $T_{\rm FL}$ below which $1/T_1T$ = constant (black dashed line) behavior is observed. 
   The insets in (a), (c), (e), and (g) show the $T$ dependence of  the ratio $R$ $\equiv$ $(1/T_1T)_{ab}$/$(1/T_1T)_{c}$. 
    The black and red horizontal lines in the insets represent the expected values for stripe-type ($R$ = 1.5) and  N\'eel-type ($R$ = 0.5) AFM spin fluctuations, respectively. }
\label{fig:Fig6}
\end{figure*}

   Similarly, Figs. \ref{fig:Fig6}(a)-\ref{fig:Fig6}(g) show the $T$ dependence of  $(1/T_1T)_{\rm ab}$ by red circles for the measured pressures at $x = 0.29$.  
   At high temperatures above $\sim$~100~K, $(1/T_1T)_{\rm ab}$ increases monotonically with increasing $T$, similar to $K_{ab}$ and $K_{c}$ for all the measured $p$ region up till 1.95 GPa. 
  As in the case of $x$ = 0.15, the deviation between $(1/T_1T)_{\rm ab}$ and $K$ is seen below $\sim$ 100 K. 
   The characteristic $T_{\rm {FL}}$ is also observed at $\sim$ 20--30 K for all $p$ values, where a constancy in $1/T_1T$ and $K$  (both $K_{\rm ab}$ and $K_{\rm c}$) is observed.

   Since the $T$ dependence of 1/$T_1T$ at low temperatures seems to be different below and above $x_{\rm c}$ and/or $p_{\rm c}$, it is important to see whether the wave vectors associated with the AFM spin fluctuations change or not. 
     According to previous NMR studies on Fe pnictides \cite{KitagawaSrFe2As2,Kitagawa2010,FukazawaKBaFe2As2} and related materials \cite{Furukawa2014,Pandey2013,Ding2016}, the ratio, $R \equiv \frac{T_{1,c}}{T_{1,ab}}$ provides valuable information on the $\bf{Q}$ dependence of the AFM spin fluctuations, where $T_{1,c}$ and $T_{1,ab}$ represent the $T_1$ measured under $H \parallel c$ and $H \parallel ab$, respectively. 
    If AFM spin fluctuations are characterized by stripe-type with $\bf{Q_{\rm 1}}$ and/or $\bf{Q_{\rm 2}}$,   an $R$ = 1.5  is expected for isotropic fluctuations while, in the case of N\'eel type AFM spin fluctuations with the wave vector $\bf{Q}$ = ($\pi$, $\pi$),  $R$ is expected to be 0.5. 
    Therefore we measured  $1/T_1T$ for $H \parallel c$ for both $x = 0.15$ and 0.29. 
The results are shown by gray circles in Figs. \ref{fig:Fig5} and Figs. \ref{fig:Fig6}, respectively where the $T$ dependence of $R$ at different pressures is shown in the corresponding inset.    
      At high temperatures above $\sim$ 100~K,  $1/T_1T$ values are nearly same for $H \parallel ab$ and $H \parallel c$ directions and $R$ $\sim$ 1 for both $x = 0.15$ and 0.29. 
    When AFM spin fluctuations start developing as $T$ is lowered below 100~K, a difference is seen for the two measured directions, and $R$ becomes $\sim$ 1.5 or more. 
   This is very similar to what was observed for $x = 0.09$ \cite{Rana2020} and suggests that AFM spin fluctuations are characterized by the stripe-type ones with  $\bf{Q_{\rm 1}}$ and/or $\bf{Q_{\rm 2}}$  in FeSe$_{1-x}$S$_{x} $ throughout the measured $p$ region up to $\sim$ 2 GPa.

\section{Discussion}

%\subsection{Evolution of AFM spin fluctuations and Phase Diagram}
\begin{figure*}%[tb]
\centering
\includegraphics[width=1.5\columnwidth]{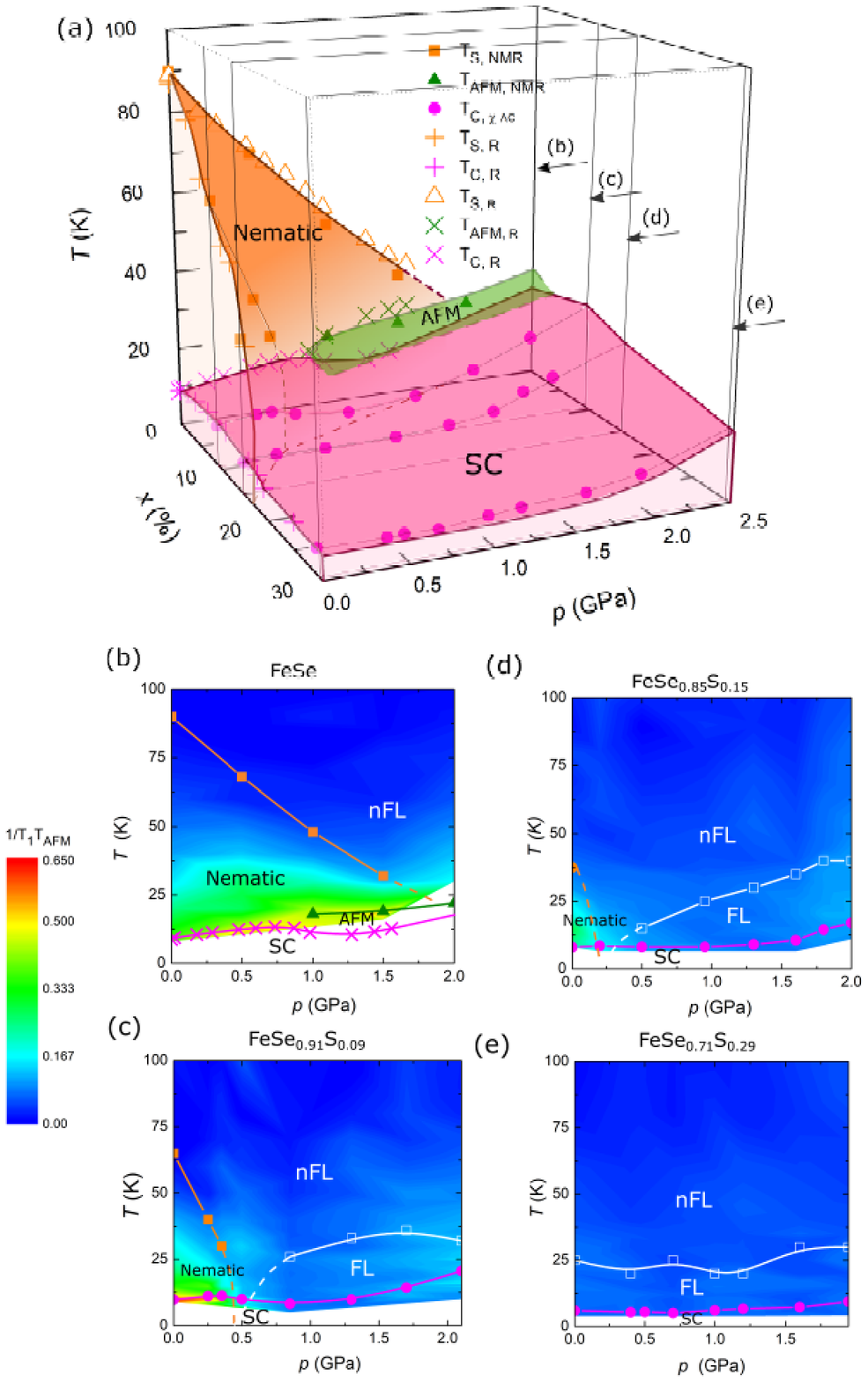} 
\caption {(a) $x$-pressure($p$)-temperature($T$) phase diagram of FeSe$_{1-x}$S$_{x} $ determined from NMR and resistivity measurements. 
The nematic transition temperatures $T_{\rm s}$ was determined by the splitting of the $^{77}$Se NMR spectra with $H \parallel ab$ in the tetragonal [110] direction which includes data from Refs.~\cite{Wiecki2017,Rana2020}. 
The SC transition temperature $T_{\rm c}$ was determined by ${\it in~situ}$ ac susceptibility (closed magneta circles) measurements at $H = 0$ \cite{Sup,Rana2020} and resistivity measurements from Ref. \cite{Udhara2016} (magenta $\times$) and Ref. \cite{Reiss2017} (magenta +). $T_{\rm N}$ for AFM transition temperature was determined by $\chi_{\rm AC}$ (closed green triangles) \cite{Wiecki2017} and resistivitiy measurements (green $\times$) \cite{Udhara2016}. (b)-(e) $p-T$ Contour plots showing the magnitude of antiferromagnetic spin fluctuations for various values of $x$ along with the transition temperatures as described in (a). 
The 1/$T_1T$ data for $x = 0$ and 0.09 are from Ref. \cite{Wiecki2017} and Ref. \cite{Rana2020}, respectively.
The Fermi liquid (FL) temperature ($T_{\rm FL}$) is the temperature at which Korringa relation was observed where both $^{77}$Se nuclear relaxation rate divided by temperature ($1/T_1T$) and $^{77}$Se Knight shift values are constant. 
  The temperature region where Korringa relation is not satisfied, is indicated as non Fermi Liquid (nFL) state. 
  The solid and dotted lines are guides for the eyes.}
\label{fig:Fig7}
\end{figure*}

   In this section, we show the relationship  between AFM spin fluctuations, nematicity and superconductivity in FeSe$_{1-x}$S$_x$.
   To discuss the magnitude of AFM spin fluctuations, we estimate the contribution of AFM spin fluctuations from the observed $1/T_1T$ assuming that the observed $1/T_1T$ can be decomposed into two components of the AFM [$(1/T_1T)_{\rm AFM}$] and the $q$ independent [$(1/T_1T)_{\rm 0}$]  components: 1/$T_1T$ = $(1/T_1T)_{\rm AFM}$  + $(1/T_1T)_0$.  
   The second term $(1/T_1T)_0$ is expected to be proportional to $K_{\rm spin}^2$. 
   Here  $K_{\rm spin}$ is given by $K - K_0$ where $K_0$ is the $T$-independent orbital shift. 
   Therefore, by comparing 1/$T_1T$ to $K_{\rm spin}^2$, one can estimate $(1/T_1T)_{\rm AFM}$. 
   The green curves in Figs. \ref{fig:Fig5} and \ref{fig:Fig6} are the estimated temperature dependence of $CK_{\rm spin}^2$; the values  are shown on right axes.
   $C$ is a proportional constant introduced to fit the temperature dependence of $(1/T_1T)_{\rm ab}$ at high temperatures above $\sim$100 K.
   From the fittings, we found  that $K_0$ = 0.175 \% is independent of $p$ and $x$ and the $p$ independent values of $C$ = 8.5 and 7.75 for  for $x$ = 0.15 and 0.29, respectively.
    Here we used $K_c$ data to compare the temperature dependence of $(1/T_1T)_{\rm ab}$ since the enhancements in $1/T_1T$ due to the stripe-type AFM spin fluctuations are  prominent along the $c$ axis \cite{Kitagawa2010}.
   For the cases where $K_c$ values were not available as in Figs. 4(b), 4(e), 4(g) and Figs. 5 (b), 5(e), 5(f), smooth extrapolations were made to estimate $K_c$  based on the data at nearest pressures for each compound. 
   Thus, the differences between $(1/T_1T)_{\rm ab}$ (red circles) and green curves are attributed to the contributions of AFM spin fluctuations, $(1/T_1T)_{\rm AFM}$.     

    The estimated AFM contributions to $1/T_1T$  are shown with the phase diagram in Figs. \ref{fig:Fig7}(b)-\ref{fig:Fig7}(d) where contour plots of $(1/T_1T)_{\rm AFM}$ are shown for $x$ = 0.15 and 0.29 as a function of $p$ together with the data for $x$ = 0 and 0.09 estimated from our previous papers \cite{Wiecki2017, Rana2020}.  
    The $p-x-T$ phase diagram of FeSe$_{1-x}$S$_x$ in Fig.~\ref{fig:Fig7}(a) was constructed from NMR measurements from ambient to $p$ $\sim$ 2 GPa  as well as the data available from literature. 
%     The  phase diagram for $x$ = 0 and 0.09 were taken from Ref. \cite{Wiecki2017} and Ref. \cite{Rana2020} respectively. 
%   The phase diagrams for $x = 0.15$ and 0.29 were constructed using the newly reported data in this paper. 
%    In the phase diagram,  the nematic transition temperatures ($T_{\rm s}$) were determined from NMR spectra where two peaks were observed  due to the twin nematic domains.  
    $T_{\rm c}$ (closed magenta circles) \cite{Sup,Rana2020} and $T_{\rm N}$ (closed green triangles) \cite{Wiecki2017} were deduced using ${\it in~situ}$ ac susceptibility measurements at zero field, and $T_{\rm s}$ (closed orange squares) \cite{Wiecki2017, Rana2020} were determined from the NMR spectra. 
For FeSe under pressure, transition temperatures for the nematic (open orange triangles), AFM (green crosses) and superconducting states (magenta crosses) determined from resistivity curves were taken from Ref.~\cite{Udhara2016}. 
   Similarly, for $p$~=~0 in FeSe$_{1-x}$S$_x$, the transition temperatures for the nematic (orange pluses) and superconducting (magenta pluses) were taken from Ref.~\cite{Reiss2017} which were also based on resistivity measurements.

    %SC$_{\rm {C2}}$ and SC$_{\rm {C4}}$ denote the SC states in the orthorhombic and tetragonal crystal structures, respectively.
   %$T_{\rm {FL}}$ is the characteristic temperature
 
   In the contour plots [Figs.~\ref{fig:Fig7}(b)-\ref{fig:Fig7}(e)], we also show $T_{\rm s}$ and $T_{\rm c}$ by  symbols in orange and magenta, respectively. 
    The white squares represent the characteristic temperature ($T_{\rm {FL}}$) below which FL states are observed.  % below which  FL state is observed. 
   It is clearly seen in the contour plots that AFM spin fluctuations are more enhanced inside the nematic state than outside the nematic state. 
   One of the interesting phenomena in FeSe$_{1-x}$S$_{x} $ is that a FL state is observed only in the C4 state after suppressing the nematic state with $p$ or $x$ throughout the presented $p-x-T$ phase diagram.
%, except at $x = 0$ where an AFM order was observed after suppressing $T_{\rm s}$ at pressures higher than $\sim$ 1 GPa. 
   A FL behavior was also seen through resistivity measurements for $x = 0.11$ under pressure \cite{Reiss2019}. 
   This was recently attributed as the signatures of quantum Griffiths phase close to the nematic quantum phase transition through magnetoresistivity measurements \cite{Reiss2021}. % and Ref.(new work).
   It is interesting to note that the smooth extrapolation of $T_{\rm FL}$  to $T_{\rm FL}$ = 0 seems to cross the nematic QPT pressure of  $p_{\rm c}$ $\sim$  0.5 and 0.2 GPa for  $x$ = 0.09 and 0.15, respectively.    
   Therefore our results also suggest the strong connection between nematic QPT and the appearance of FL states. 
%   It is also seen that SC$_{\rm {C4}}$ always arises from a FL state.  %Exceptions to this observation are  SC$_{\rm {C4}}$  states observed near $p_{\rm c}$, for example at $(x,p)$ values of (0.9, 0.5 GPa) and (0.15, 0.2 GPa), when the crossing from $SC_{C2}$ to $SC_{C4}$ occurs.

% \subsection{Role of nematicity on the relationship between $T_{\rm c}$ and Antiferromagnetic spin Fluctuations}
%\label{D.5.5.3}
\begin{figure}[h!tb]
\centering
\includegraphics[width=1\columnwidth]{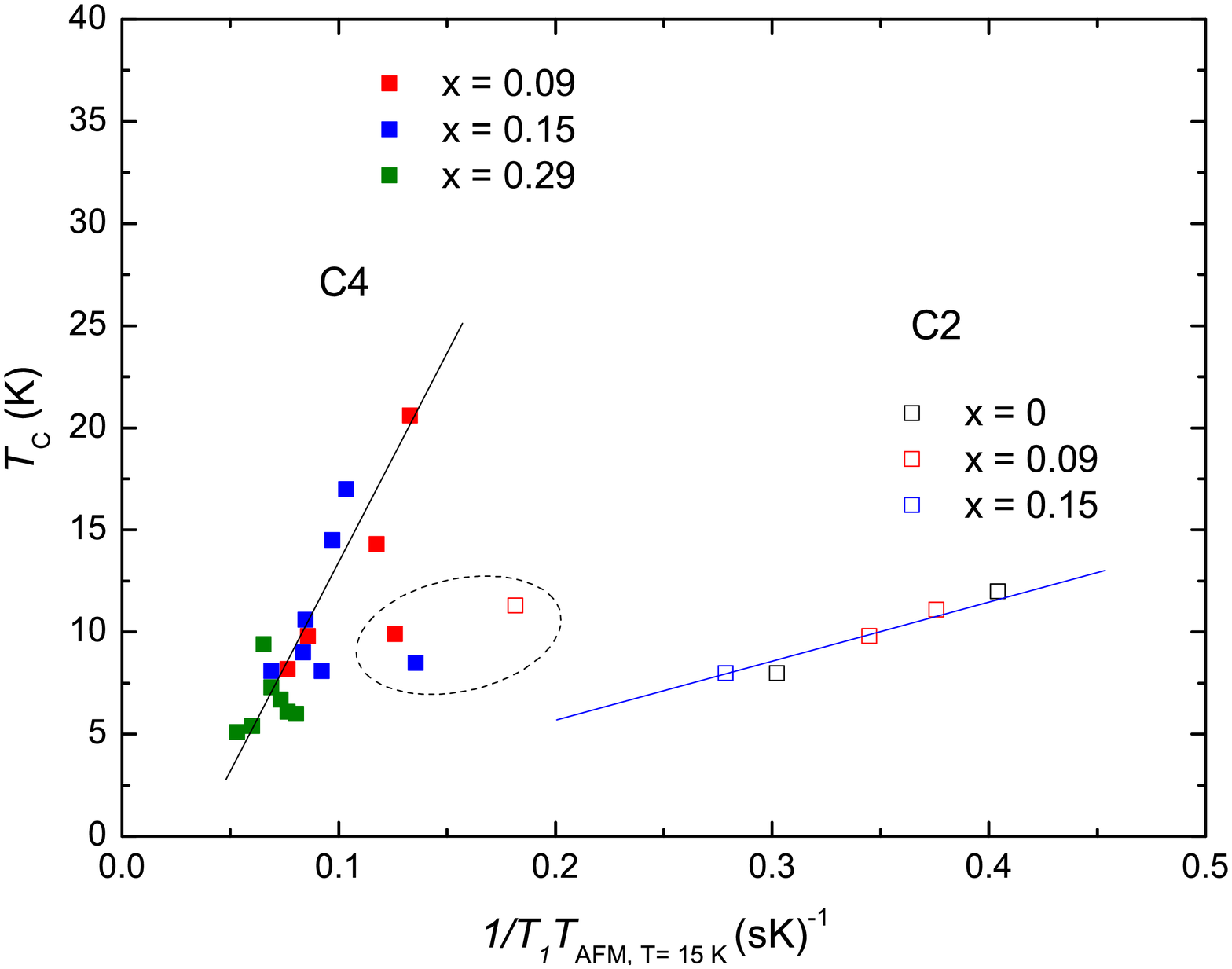} 
\caption{Plot of $T_{\rm c}$ at zero field versus $(1/T_1T)_{\rm AFM}$ at $T$~=~15~K. 
The values for $x = 0$ are taken from Refs \cite{Wiecki2017,Bohmer2015,Udhara2016} and for $x = 0.09$  from Ref. \cite{Rana2020}. 
   Black and blue lines show linear fits for  C4 (closed symbols) and C2 (open symbols) phases, respectively. 
  The three points inside the dashed black oval denote the values taken close to the critical pressure for nematic quantum phase transition and were not included in the linear fits.}
\label{fig:DFig9}
\end{figure}

\begin{figure}%[tb]
\centering
\includegraphics[width=1\columnwidth]{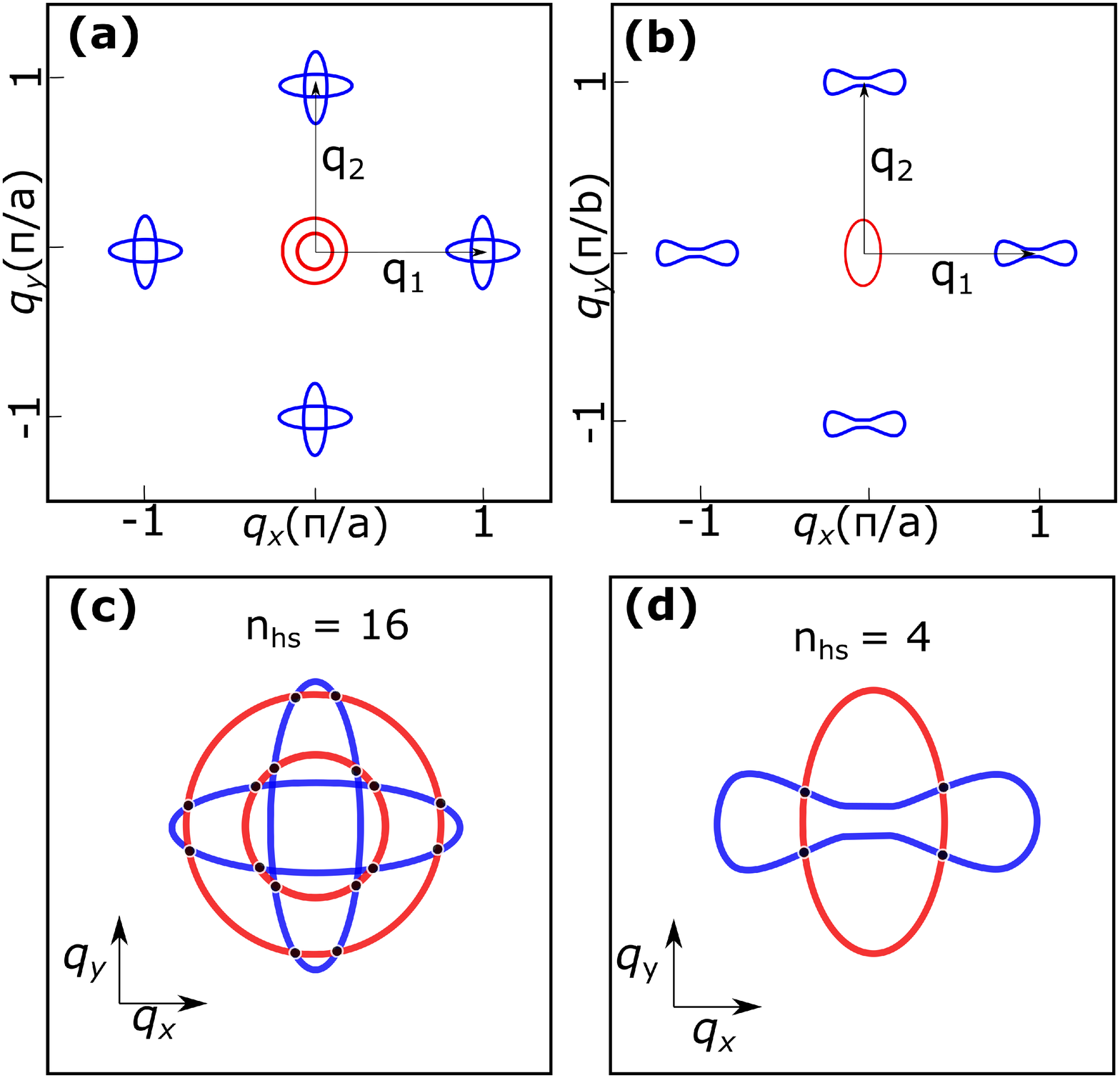} 
\caption[Schematics Fermi surfaces and nesting with stripe type wave vector at $q_z = 0$ in the FeSe$_{1-x}$S$_{x} $ system.]{Schematic Fermi surfaces and nesting with stripe-type wave vector at $q_z = 0$ in the FeSe$_{1-x}$S$_{x} $ system. 
(a) Schematic Fermi surface for the tetragonal state based on Refs. \cite{Reiss2017,Watson2017} where $q_x$ and $q_y$ are in units of $\pi/a$ where a is related to the tetragonal lattice constant $d$, by $a\equiv d/\sqrt{2}$. 
(b) Schematic Fermi surface for the detwinned nematic state based on Ref. \cite{Watson2017} where $q_x$ and $q_y$ are in units of $\pi/a$ and $\pi/b$ where $a$ and $b$ are the orthorhombic lattice constants. 
  Here red lines represent the hole pockets at the $\Gamma$ point and blue lines represent the electron pockets at the Z and A points for the tetragonal and orthorhombic states respectively. 
(c-d) The number of hotspots ($n_{\rm {hs}}$) due to the intersection of single hole and single electron pockets when either is translated by the wave vectors $\bf{Q_{\rm 1}}$ or $\bf{Q_{\rm 2}}$ in (c) tetragonal and (d) nematic states.}
\label{fig:DFig10}
\end{figure}

   Now we discuss the relationship between $T_{\rm c}$ and the magnitude of AFM spin fluctuations.
   In Fig. \ref{fig:DFig9}, we plot  $T_{\rm c}$  at zero magnetic field against the magnitude of AFM spin fluctuations. 
    Here we used the values of (1/$T_1T)_{\rm AFM}$ at $T$ = 15 K to represent the magnitude of AFM spin fluctuations.
   As has been pointed out in Ref. \cite{Rana2020}, the data points are mainly divided into two groups with the open (C2 state) and closed (C4 state) squares, and
 $T_{\rm c}$ seems to be proportional to AFM spin fluctuations in both C2 and C4 states as shown in the blue and black lines, respectively. 
    However, clearly the slope for the C4 states is higher that that for the C2 state.
    This indicates that the AFM spin fluctuations without nematicity enhance $T_{\rm c}$ much more than those with nematicity. 
    Here, the black line has a ~ $7 \pm 2$ times higher slope than the blue line.
% , signifying that the C4 symmetric AFM spin fluctuations are more favorable in enhancing superconductivity. 
    It is interesting to point out that the three points inside the black dashed oval between the two lines correspond to the data close to $p_{\rm c}$,  suggesting a smooth crossover between the two linear behaviors at the nematic QPT.

    One of the reasons why AFM spin fluctuations in the C4 state lead to higher $T_{\rm c}$ compared to those in the C2 state  in FeSe$_{1-x}$S$_{x} $ could be due to the different Fermi surfaces in the  paramagnetic tetragonal and  nematic orthorhombic phases.
   Figures \ref{fig:DFig10}(a) and \ref{fig:DFig10}(b) show the schematic Fermi surfaces at $q_z$ = 0 for the C4 and C2 states based on Refs. \cite{Reiss2017,Watson2017}. 
   The main differences are seen at $\bf{q}=(q_x,q_y)=(0,0)$  ($\Gamma$ point) and at $\bf{q}=\pm \bf{Q_{\rm 1}}$ and 
$\bf{q}=\pm \bf{Q_{\rm 2}}$ (Z points in the C4 state and A points in the C2 state). 
   For simplification, the orbital contribution to the Fermi surface are not indicated.

    In the C4 state of FeSe, two circular hole pockets (red lines) at $\Gamma$ point and two oval shaped electron pockets (blue lines) at the Z points are found in ARPES measurements \cite{Watson2017}. 
   This structure of the Fermi surface is also observed  in ARPES measurements with $x = 0.18$ at the low $T$ of 10 K where nematicity is absent \cite{Reiss2017}. 
   When the hole pocket at $\Gamma$ point is translated by either $\bf{Q_{\rm 1}}$ or $\bf{Q_{\rm 2}}$, 16 intersections with the electron pockets at the Z points can exist as shown in Fig. \ref{fig:DFig10}(c). 
   Here $n_{\rm hs} = 16$ denotes the number of intersections when the hole pockets are translated by one wave vector. 
   Since the four wave vectors ($\pm \bf{Q_{\rm 1}}$ and $\pm \bf{Q_{\rm 1}}$) are allowed for C4 symmetry,  a total of $16 \times 4 = 64$ possible intersections can be found. 

   In the C2 nematic state,  on the other hand, an oval hole pocket (red line) at the $\Gamma$ point and a single peanut shaped electron pocket  (blue line)  at  the A points are present as shown in Fig. \ref{fig:DFig10}(b) \cite{Watson2017}. 
    When the hole pocket at $\Gamma$ is translated by either $\bf{Q_{\rm 1}}$ or $\bf{Q_{\rm 2}}$ to the electron pocket at A points in the nematic state, there are 4 possible intersections (that is,  $n_{\rm hs}$ = 4) for one wave vector as shown in Fig.~\ref{fig:DFig10}(d). 
   Since two wave vectors (either $\pm \bf{Q_{\rm 1}}$ or $\pm \bf{Q_{\rm 1}}$) are prominent in the nematic state, a total of $4 \times 2 = 8$ intersections can be present.   % with  intersect at 4 points to the electron pockets in A site for the nesting condition of +q, and 4 more points for -q. 

   These intersections that meet the nesting conditions are called hotspots and are deemed important in determining $T_{\rm c}$ in high-$T_{\rm c}$ cuprates where AFM spin fluctuations are relevant \cite{Wang2017}. 
   The total number of the hotspots  in the C4 symmetric  tetragonal state  increases by a factor of 64/8=8 with respect to  the C2 orthorhomibc nematic state.
  This number seems to be consistent with the experimentally determined factor of $\sim$~$7\pm2$  in the slopes of the lines of Fig. \ref{fig:DFig9}.
   Although this is quite qualitative, our results suggest that \lq\lq{}hotspots\rq\rq{} are one of the key parameters to determine $T_{\rm c}$ in Fe-based superconductors.
    More detailed theoretical studies for quantitative analysis about  the dependency of $T_{\rm c}$ on $n_{\rm hs}$ are highly called for.

\begin{figure}[h!tb]
\centering
\includegraphics[width=\columnwidth]{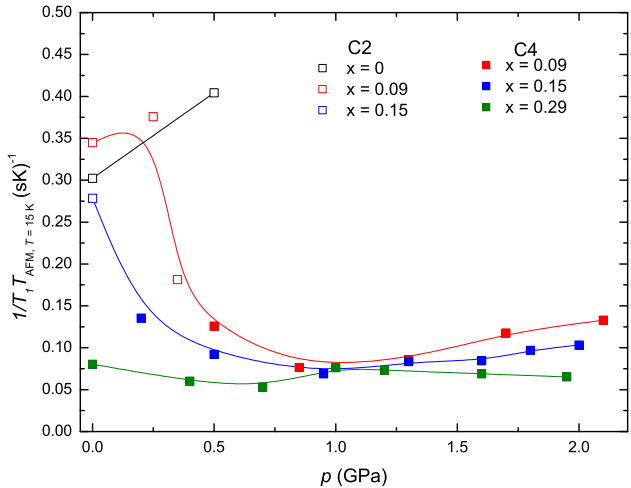} 
\caption[Pressure dependence of $(1/T_1T)_{\rm AFM}$ at the temperature of 15 K in FeSe$_{1-x}$S$_{x}$]{Pressure dependence of $(1/T_1T)_{\rm AFM}$ in FeSe$_{1-x}$S$_{x}$.  Here the values for $x = 0$ and $x = 0.09$ are taken from Refs. \cite{Wiecki2017,Bohmer2015} and Ref. \cite{Rana2020},  respectively.}
\label{fig:DFig8}
\end{figure}   

%\red{Not sure if we include}
   Finally we comment on  the $x$ and $p$ dependences of AFM correlation length $\xi_{\rm AFM}$.
    Figure \ref{fig:DFig8} shows the  $x$ and $p$ dependent of $(1/T_1T)_{\rm {AFM}}$ at $T$ = 15 K in FeSe$_{1-x}$S$_{x}$ where the open and closed symbols represent the C2 and C4 states, respectively. 
    It is clearly seen that the values of $(1/T_1T)_{\rm {AFM}}$ in the C2 phase are higher than those in the C4 phase. 
    According to Millis-Monien-Pines (MMP) model,  $(1/T_1T)_{\rm AFM}$ is proportional to the square of $\xi_{\rm AFM}$ in real space ($1/T_1T \propto \xi_{\rm AFM}^2$) \cite{Millis1990,Baek2020}. 
   Therefore our results may suggest the $\xi_{\rm AFM}$ in the C4 state is shorter than in the nematic C2 state.
   It is interesting to point out that SC states with nematicity (C2 state) and without nematicity (C4 state) are induced by AFM spin fluctuations with different values of $\xi_{\rm AFM}$.
   It would also be interesting if the change in $\xi_{\rm AFM}$ between the C4 and C2 states relate to the BCS-BEC transition at $x_{\rm c}$ $\sim$ 0.17 observed by ARPES measurements at ambient pressure \cite{Hashimoto2021} where longer $\xi_{\rm AFM}$ is suggested when BCS superconductivity is present and a shorter $\xi_{\rm AFM}$ is suggested in the BEC case. 
  If so, the shortening of $\xi_{\rm AFM}$ under $p$ in FeSe$_{1-x}$S$_{x} $ would be compatible to the BCS-BEC transition at the respective critical pressure for nematic QPT around 0.5 GPa for $x~=~0.09$ and 0.2 GPa for $x~=~0.15$. Further studies are called for searching BCS-BEC transition under pressure in FeSe$_{1-x}$S$_{x}$.

\section{Summary}
%\red{NOT YET} 

In summary, $^{77}$Se-NMR measurements were carried out in the FeSe$_{1-x}$S$_{x} $ system under pressure for $x$ = 0.15 and 0.29, and were analyzed together with the previously reported measurements for $x$ = 0 \cite{Wiecki2017} and 0.09 \cite{Rana2020}, to provide a comprehensive study of the interrelationships between nematicity, magnetism and superconductivity in this system. % found at the  Bardeen-Cooper-Schrieffer to Bose-Einstein-condensate crossover. % understand the relationship between the properties of antiferromagnetic fluctuations and superconductivity. 
We established  the $p-x-T$ phase diagram featuring the presence of nematic quantum phase transitions, the appearance of Fermi liquid phases, superconductivity with C4 and C2 rotational symmetries and the contour plots of the magnitude of AFM spin fluctuations. %The AFM spin fluctuations were tuned using $p$ by suppressing the nematic order for samples with $x$ values lower than the critical sulfur level, $x_{\rm C}~\sim~0.17$, for nematic quantum phase transition. 
% A continuous shortening of AFM correlation length occurs at the nematic quantum phase transition regions of the $x\mhyphen p\mhyphen T$ phase diagram. 
   When analyzing the dependence of superconducting transition temperature on the magnitude of AFM spin fluctuations, the different linear relationships on the basis of symmetry are apparent. 
   The C4 symmetric AFM spin fluctuations are more effective in enhancing $T_{\rm c}$ by a factor of 7~$\pm$~2 higher than compared to their C2 symmetric counterparts. 
  This was explained in terms of the presence of higher total number of hotspots in the C4 symmetric Fermi surface compared to the Fermi surface with C2 symmetry in the FeSe$_{1-x}$S$_{x} $ system by a factor of 8.  
  These results indicate that the hotspots may play an important role in determining  $T_{\rm c}$ in  FeSe$_{1-x}$S$_{x} $. 
  Future theoretical investigations will be interesting to clarify  the relationship between the total number of hotspots and $T_{\rm c}$, leading a step closer towards understanding the details of Cooper-pairing mechanism in unconventional superconductors.

\section{Acknowledgments}
We appreciate Rafael Fernandes for helpful discussions.
The research was supported by the U.S. Department of Energy (DOE), Office of Basic Energy Sciences, Division of Materials Sciences and Engineering. Ames Laboratory is operated for the U.S. DOE by Iowa State University under Contract No.~DE-AC02-07CH11358.

\end{document}